\documentclass{mem}
\usepackage{natbib}\usepackage{txfonts}\usepackage{balance}
\usepackage{graphicx}
\usepackage[a4paper,breaklinks,dvipdfm]{hyperref}
\idline{75}{282}
\begin{document}
\def\teff{$T\rm_{eff }$}
\def\kms{$\mathrm {km s}^{-1}$}

\title{
Chemical abundances of blue straggler stars
in Galactic Globular Clusters
}

   \subtitle{}

\author{
L. \,Lovisi\inst{1} 
          }

  \offprints{L. Lovisi}

\institute{
Dipartimento di Fisica e Astronomia, Universit\`a degli Studi
di Bologna, Viale Berti Pichat 6/2, I--40127 Bologna, Italy\\
\email{loredana.lovisi2@unibo.it}
}

\authorrunning{Lovisi}

\titlerunning{Chemical abundances of blue straggler stars in Galactic Globular Clusters}

\abstract{
By using the high resolution spectrograph FLAMES@VLT we performed the first systematic 
campaign devoted to measure chemical abundances of blue straggler stars (BSSs). These stars, whose existence is not predicted by 
the canonical stellar evolutionary theory, are likely the product of the interactions between stars in the dense 
environment of Globular Clusters. Two main scenarios for BSS formation (mass transfer in binary 
systems and stellar collisions) have been proposed and hydrodynamical simulations predict different chemical patterns in the two cases, in
particular C and O depletion for mass transfer BSSs. In this contribution, the main results for BSS samples in 6 Globular Clusters 
and their interpretation in terms of BSS formation processes are discussed. For the first time, evidence of radiative levitation in the shallow
envelopes of BSSs hotter than $\sim$8000 K has been found. C and O depletion for some BSSs has been detected in 47 Tucanae, M30 and $\omega$ Centauri 
thus suggesting a mass transfer origin for these BSSs. The low percentage of CO-depleted BSSs suggests that 
depletion might be a transient phenomenon.

}
\maketitle{}

\section{Introduction}

Dense Globular Clusters (GCs) are efficient ``furnaces'' for generating exotic objects
such as low-mass X-ray binaries, cataclysmic variables, millisecond pulsars and blue straggler stars (BSSs). Most of these
objects are thought to be formed by stellar interactions between single or (more likely) binary stars. In particular,
among these exotica, BSSs are the most numerous. 
Firstly discovered by Sandage in the GC M3 \citep{Sandage53}, they are brighter and bluer than the turnoff and they are more massive 
than normal MS stars \citep{Shara97, Fiorentino13}, thus mimicking a rejuvenated stellar population. According to the canonical
stellar evolutionary theory, 
they should have already exhausted their nuclear fuel and evolved in cool white dwarfs long time ago. Moreover, since 
they are the most massive stars in GCs (and thus they are strongly affected by dynamical friction), 
they are ideal tools to probe the dynamical evolution of the hosting cluster. Indeed, recent results by \citet{Ferraro12} 
show that BSS radial distribution in GCs can be used as a \textit{dynamical clock} that provides a direct measurement of the cluster dynamical age.\\
In spite of their importance, BSS formation mechanisms and their main properties are very poorly known.
Two main scenarios have been proposed for their formation: BSSs might originate through mass transfer (MT) in binary systems
or through stellar collisions (COL), that are expected to be frequent in the dense cores of GCs. MT- and COL-BSSs are expected to have
different chemical patterns. In fact, according to hydrodynamical simulations, MT-BSSs should show 
C and O depletion \citep{Sarna96}, since the material should come from deep in the donor star, where the CNO cycle
already occurred. On the contrary, negligible mixing between inner cores and outer envelopes is expected for COL-BSSs that should
show normal C and O abundances \citep{Lombardi95}. \\
In order to deeply investigate the chemical properties of these puzzling objects, in 2006 we started a systematic campaign
devoted to secure high-resolution spectra with FLAMES@VLT for BSSs in 6 GCs, namely 47 Tucanae, M4, NGC 6752, NGC 6397, 
M30 and $\omega$ Centauri. The selected clusters span a wide range of properties, in particular metallicity 
and dynamical status. Moreover, the sample includes two peculiar GCs, namely M30 and $\omega$ Centauri. Indeed,
M30 is known to host two distinct and well separated BSS sequences \citep{Ferraro09}, whereas almost all the BSS population in $\omega$ Centauri
is though to have non-collisional origin (and thus to be formed mainly through MT, \citealt{Ferraro06a}).
\section{47 Tucanae: first detection of CO-depletion}
First results have been obtained in 47 Tucanae \citep{Ferraro06b}: the analysis of 42 BSSs led to the discovery of 6 stars with
low C and O abundances. These abundances disagree with C and O content of the other BSSs in the sample and also of cluster MS stars. 
Moreover, such a depletion is totally incompatible with the second generation scenario, according to which second generation stars (depleted 
in C, O and other light elements) have been formed from material reprocessed by the first generation stars. On the contrary, 
low C and O abundances for these 6 BSSs can be interpreted in terms of the MT process.
\section{M4: the largest percentage of fast rotating BSSs}
A different result has been obtained for M4 \citep{Lovisi10} where all the studied BSSs share the same chemical composition of MS stars and no
CO-depleted BSSs have been found. According to the percentage (14\%) identified in 47 Tucanae, 1-2 CO-depleted
BSSs should have been expected in M4. However, the most intriguing result in the study of this sample does not
concern chemistry but rotational velocities. In fact, a large percentage (40\%) of fast rotating BSSs (rotational velocities higher than
40\kms, up to more than 100\kms) has been found. This is the first time that such a high fraction of fast rotating BSSs has 
been identified. However, due to their high rotational velocity, for 8 out of 20 BSSs it was not possible to derive C and O abundances.
\section{NGC 6752 and NGC 6397: the occurrence of radiative levitation}
\begin{figure}[]
\resizebox{\hsize}{!}{\includegraphics[clip=true]{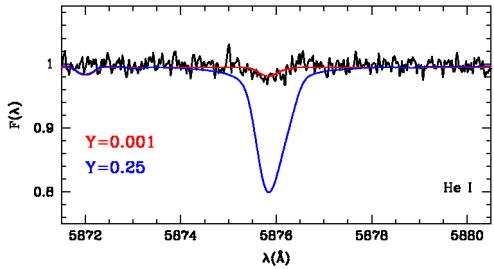}}
\caption{HeI line at $\sim$5876 \AA\ for a BSS of NGC 6397.\footnotesize}
\label{he}
\end{figure}

Also the study of NGC 6752 \citep{Lovisi13a} reveals the lack of CO-depleted BSSs and their total compatibility with C and O abundances of dwarf stars. 
Nevertheless, for these BSSs the iron distribution is quite spread ($\sim$1 dex) and shows a trend with the stellar
temperature, with higher iron content for hotter stars. The same trend was already observed by \citet{Lovisi12}
in NGC 6397. In this cluster, BSSs show an even larger spread in the iron content (up to more than 2 dex) and some BSSs have solar or
super-solar values. This behavior is known to occur in the Horizontal Branch in stars hotter than $\sim$11000 K that have 
high iron contents up to solar values \citep{Behr03}. The common interpretation for this feature is the occurrence of radiative levitation
in the atmospheres of stars with shallow or non-convective envelopes, when the radiative acceleration overcomes the 
gravity one. The main effect is that many chemical elements are pushed toward the stellar surface, thus altering the initial
chemical composition. Another diffusion process generally occurring simultaneously with radiative acceleration is the gravitational
settling of He. For one BSS of NGC 6397, we secured also high signal-to-noise UVES spectra with wavelength coverage suitable 
to observe the HeI line at $\sim$5876 \AA. A comparison with synthetic spectra computed with the same atmospheric parameters, metallicity 
and rotational velocity of the star, reveals that the observed He line cannot be fitted with a normal He mass fraction
(Y=0.25) and that a much lower He content (Y=0.001) is needed, thus confirming the occurrence of diffusion processes (see Fig. \ref{he}).

\section{Two peculiar clusters: M30 and \\
$\omega$ Centauri}
High quality and accurate color-magnitude diagrams of M30, revealed the presence of two parallel BSS sequences, similarly populated
and very well separated. According to the interpretation by \citet{Ferraro09}, the bluer sequence is made by BSSs formed through collisions
whereas the redder one hosts mainly BSSs formed through MT. Since the two sequences are so well separated, a short lived event (instead of
a continuous process) should have occurred 1 or 2 Gyr ago and boosted the BSS formation through both the MT- and COL-process. The intriguing
presence of two BSS sequences makes M30 the ideal target for studying the difference in the chemical properties of MT- and COL-BSSs and
searching for CO-depletion. Nevertheless, observing BSSs in M30 is a quite challenging task, since they are relatively faint (V magnitude between
17 and 18.8) and very concentrated toward the cluster center. Moreover, the brightest BSSs are also the hottest ones and therefore likely affected by
radiative levitation. In spite of these difficulties, we obtained 11 spectra including 8 BSSs in the red sequence and 3 in the blue one. 
As expected, most of the stars suffer from radiative levitation and thus the chemical analysis has been performed only on the 5 coldest BSSs \citep{Lovisi13b}. 
Unfortunately they all belong to the red sequence, thus preventing the comparison with BSSs in the blue one. For these stars, only O upper 
limits have been obtained (whereas no information of C abundances could have been inferred). O upper limits are lower than 0.3 dex and 
incompatible with abundances of RGB stars in the same cluster (therefore a second generation origin for these stars should be discarded).\\
Concerning $\omega$ Centauri, it is one of the most puzzling stellar systems in our Galaxy. The flat radial distribution of its BSS population \citep{Ferraro06a}
suggests that this cluster is not dynamically relaxed, since dynamical friction did not have enough time to occur yet. For this reason, stellar
interactions in this cluster are thought to be very rare and almost the entire BSS population is probably formed through MT in
primordial binary systems. Therefore, $\omega$ Centauri is the best target for investigating chemical properties of MT-BSSs.
Despite the large intrinsic spread in its iron content (due probably to a non-genuine origin of $\omega$ Centauri), the analysis of 
the [Fe/H] ratio as a function of stellar temperatures reveals the occurrence of radiative levitation also in the
hottest stars of this cluster. In order to avoid alteration in the chemical abundances due to radiative levitation effects, we measured 
C and O abundances only for the 9 coldest BSSs. A comparison with results by \citet{Marino12} 
shows that most of the BSSs share the same chemical content of dwarf stars (Mucciarelli et al. 2014, in prep.). However, one BSS has 
low C abundances whereas another BSS show [O/Fe] ratio incompatible with the dwarf stars. Also in this case, an interpretation in terms 
of the occurrence of a MT process has been provided.

\section{Conclusions}
\begin{figure}[]
\resizebox{\hsize}{!}{\includegraphics[clip=true]{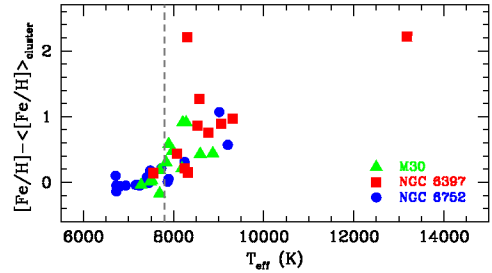}}
\caption{[Fe/H] ratios (scaled on the mean cluster iron content) as a function of stellar temperatures for the BSS samples of M30, NGC 6397 and NGC 6752.
\footnotesize
}
\label{lev}
\end{figure}

We performed the first systematic campaign in order to deeply investigate the chemical properties of BSSs in 6 different GCs. Our study
is based on the analysis of high-resolution spectra obtained with FLAMES@VLT. For the first time, we found evidence of the occurrence of
radiative levitation in the shallow convective envelopes of BSSs. This radiative process is usually observed in Horizontal Branch stars hotter
than 11000 K. At variance with Horizontal Branch stars, we found that radiative levitation in BSS atmospheres occurs at lower temperature
around $\sim$8000 K. This is particularly evident from Fig. \ref{lev}, where [Fe/H] ratios for the BSS samples of M30, NGC 6397 and NGC 6752 
scaled to the mean cluster iron content have been plotted.\\ 
The analysis of BSSs in the red sequence of M30, revealed that 4 out 5 BSSs (with chemical abundances not altered by radiative levitation)
are O depleted. Since these stars should have been formed through MT, this suggests that depletion might be considered a real signature of 
the MT process. However, the low percentage of CO-depleted BSSs detected in the sample (spanning from 0\% in M4 and NGC 6752 to $\sim$14\% in 47 Tucanae) 
suggests that CO-depletion might be a transient phenomenon. As already suggested by \citet{Ferraro06b}, BSSs could experience different 
evolutionary stages that might end with mixing processes restoring the original chemical abundances and thus erasing the MT signature.\\
 
\begin{acknowledgements}
This research is part of the project COSMIC-LAB funded by the
European Research Council (under contract ERC-2010-AdG-267675).
\end{acknowledgements}

\bibliographystyle{aa}

\end{document}